\documentclass[twocolumn,prx,aps]{revtex4} 
\usepackage{epsfig,amsmath}
\usepackage{graphicx}
\usepackage{dcolumn}
\usepackage{bm}
\usepackage{bbold}
\usepackage{braket}
\newcommand{\beq}{\begin{equation}}
\newcommand{\eeq}{\end{equation}}
\newcommand{\beqa}{\begin{eqnarray}}
\newcommand{\eeqa}{\end{eqnarray}}

\begin{document}
\author{Yi Ding}
\email{yding2021@swjtu.edu.cn}
\affiliation{School of Physical Science and Technology, Southwest Jiaotong University, Chengdu 610031, China}
\title{Rotationally symmetric momentum flow produced by scattering on an anisotropic random medium}
\date{\today}

\begin{abstract}
As is well known that the distribution of the scattered radiation generated by an anisotropic scatterer usually lacks rotational symmetry about the direction of incidence due to the spatial anisotropy of the scatterer itself. Here we show that the rotationally symmetric distribution of the far-zone scattered momentum flow may be realized provided that the structural parameters of both the medium and the source are chosen suitably, when a polychromatic electromagnetic plane wave is scattered by an anisotropic Gaussian Schell-model medium. We derive necessary and sufficient conditions for producing such a symmetric distribution, and further elucidated the relationship between the spectral degree of polarization of the incident source and the rotationally symmetric momentum flow of the scattered field in the far zone. It is found that the realization of the rotationally symmetric scattered momentum flow is independent of the spectral degree of polarization of the source, i.e., the rotationally symmetric distribution of the far-zone scattered momentum flow is always realizable regardless of whether the incident source is fully polarized, partially polarized or completely unpolarized. Our results may find useful application in optical micromanipulation, especially when the optical force used to manipulate particles requires to be rotationally symmetric.
\end{abstract}

\maketitle
\section{Introduction}
As an important generalization of traditional isotropic scatterer, the spatially anisotropic scatterer in the past decade has attracted considerable interest in both the scientific and engineering communities due to its elegant ability to mimic many practical scatterers, such as ellipsoids. The first introduction of anisotropic scatterer model into the classical theory of potential scattering should attribute to Du and Zhao in the study of light scattering from a Gaussian-correlated, quasi-homogeneous, anisotropic medium \cite{DXZ}. Shortly afterwards, they extended this continuous scatterer model to a collection of anisotropic particles that have deterministic distributions \cite{DXZD}. Since then, the statistical properties of light waves on scattering from different anisotropic media have been discussed extensively (see, for example, Refs. \cite{XinyueDu, X11, X22, Li11, CXinyu, LWC, LICW, L111, XLJi, XiaoningPan, Peng} and references therein). These results have shown that the far-zone distribution of the scattered radiation generated by an anisotropic medium is rotationally asymmetric about the direction of incidence \cite{DXZ, DXZD, LICW, XLJi, XiaoningPan}. Naturally, an interesting problem arises: Is it possible to obtain a rotationally symmetric scattered radiation about the direction of incidence when a light wave is scattered by an anisotopic scatterer?

In fact, a similar question was first considered by Li and Wolf in the problem of light radiation from a planar, Gaussian, Schell-model source of any state of coherence source \cite{LW}. The authors stated that with a suitable choice of the source parameters the radiant intensity may be rotationally symmetric about the normal to the source plane. Later, the question itself above was examined by Du and Zhao in the study of rotationally symmetric scattering from Gaussian-correlated, quasihomogeneous, anisotropic media \cite{DZ}. The authors reported the rotationally symmetric spectral density and spectral degree of coherence of the far-zone scattered field, and derived the necessary and sufficient conditions for producing these symmetric distributions. The aforementioned two studies are of importance, however, the results in them are confined to the scalar analysis of optical field.

In this work, within the framework of electromagnetic scattering, we are devoted to the question above by considering a rotationally symmetric momentum flow produced by scattering on an anisotopic random medium. We will first derive the tensor form of the analytic expression for the momentum flow of the far field generated by scattering of a polychromatic electromagnetic plane wave on an anisotropic, Gaussian, Schell-model medium. Based on this, we will formulate necessary and sufficient conditions for producing a rotationally symmetric scattered momentum flow in the far zone, and further elucidate the relation between the spectral degree of polarization of the incident source and the rotationally symmetric scattered momentum flow. Finally, some numerical examples will be presented to confirm our results.

The whole paper is organized as follows: we will derive the tensor form of the analytic expression for the momentum flow of a polychromatic electromagnetic plane wave on scattering from an anisotropic, Gaussian, Schell-model medium in Sec. II; The necessary and sufficient conditions for producing the rotationally symmetric distribution of the scattered momentum flow in the far zone will be presented in Sec. III. Meanwhile, some numerical examples will also be given to confirm our results in this section; The paper is summarized and the potential application of our study is prospected in Sec. IV.

\section{The tensor expression for the far-zone momentum flow generated by scattering on an random anisotropic medium}
Let us consider scattering of a polychromatic electromagnetic plane wave incident upon a linear scatterer with a finite volume $D$, as shown in Fig. 1. The incident field at a point, specified by a vector $\mathbf{r}^{\prime}$, is represented by a statistical ensemble $\bigl\{E_{i}(\mathbf{r}^{\prime},\omega)\bigr\}$ $(i=x,y,z)$ of monochromatic realizations oscillating at the frequency $\omega$. Here we set
\begin{subequations}\label{12}
\begin{align}
    E_{x}(\mathbf{r}^{\prime},\omega)&=a_{x}(\omega)e^{ik\mathbf{s}_{0}\cdot\mathbf{r}^{\prime}}, \\ E_{y}(\mathbf{r}^{\prime},\omega)&=a_{y}(\omega)e^{ik\mathbf{s}_{0}\cdot\mathbf{r}^{\prime}}, \\  E_{z}(\mathbf{r}^{\prime},\omega)&=0,
\end{align}
\end{subequations}
where $a_{i}(\omega)$ is a random amplitude of the electric field along the $i$-th axis, and $k$ is the wave number of light in vacuum. $\mathbf{s}_{0}=[0,0,1]$ is a real unit vector, which denotes the direction of incidence.
\begin{figure}[ht]
\centering
\includegraphics[width=7cm]{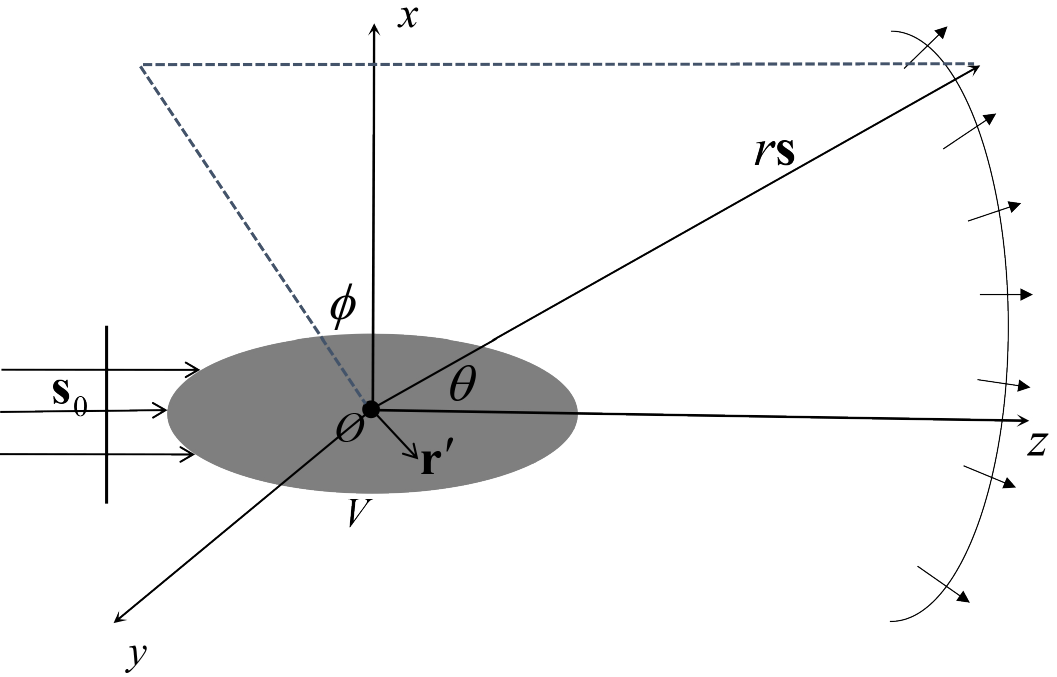}
\caption{Illustration of notations.}
\label{Fig 1}
\end{figure}

The second-order coherence properties of the incident wave at a point $\mathbf{r}^{\prime}$ can be found from its cross-spectral density matrix by setting two spatial position variables coincide, with the help of Eq. (\ref{12}), which can be formulated as
\begin{align}
\mathbf{W}(\mathbf{r}^{\prime},\mathbf{r}^{\prime},\omega)&=\left[
\begin{array}{ccc}
  S_{xx}(\omega)   &  S_{xy}(\omega)&0\\
  S_{yx}(\omega)  &  S_{yy}(\omega) &0\\
  0&0&0
\end{array}
\right],
\end{align}
where the diagonal elements represent the spectral density of the incident field, while the off-diagonal elements denote the spectral correlation between the two mutually orthogonal components $E_{x}$ and $E_{y}$ of the electric field.

Assume now that the scatterer is weak so that the scattering can be addressed within the accuracy of the first-order Born approximation \cite{Wolf1, Wolf2}. As we know, the scattered field behaves globally like a spherical wave in the far zone, which implies that the scattered field will be greatly simplified if one formulates it in the spherical polar system. Thus the scattered electric field and the scattered magnetic field at a point $r\mathbf{s}$ ($\mathbf{s}=[s_{x},s_{y},s_{z}$]) in the far zone can be separately expressed as \cite{TON}
\begin{equation}\label{ele}
\left[
\begin{array}{ccc}
   E^{(s)}_{r} \\
   E^{(s)}_{\theta}\\
   E^{(s)}_{\phi}  
\end{array}
\right]^{\top}=\frac{e^{ikr}}{r}\int_{\mathcal{V}}F(\mathbf{r}^{\prime},\omega)\mathbf{E}(\mathbf{r}^{\prime},\omega)\mathbf{A}_{1}(\theta,\phi)e^{-ik\mathbf{s}\cdot\mathbf{r}^{\prime}}d^{3}r^{\prime},
\end{equation}
\begin{equation}\label{mag}
\left[
\begin{array}{ccc}
   B^{(s)}_{r} \\
   B^{(s)}_{\theta} \\
   B^{(s)}_{\phi}
\end{array}
\right]^{\top}=\frac{e^{ikr}}{r}\int_{\mathcal{V}}F(\mathbf{r}^{\prime},\omega)\mathbf{E}(\mathbf{r}^{\prime},\omega)\mathbf{A}_{2}(\theta,\phi)e^{-ik\mathbf{s}\cdot\mathbf{r}^{\prime}}d^{3}r^{\prime},
\end{equation}
where  
\begin{equation}
\mathbf{A}_{1}(\theta,\phi)=\left[
\begin{array}{ccc}
  0 &  \cos\theta\cos\phi& -\sin\phi\\
  0  & \cos\theta\sin\phi& \cos\phi \\
  0 & -\sin\theta & 0
\end{array}
\right]
\end{equation}
and 
\begin{equation}
\mathbf{A}_{2}(\theta,\phi)=\left[
\begin{array}{ccc}
  0 &  \sin\phi& \cos\theta\cos\phi\\
  0  & -\cos\phi& \cos\theta\sin\phi \\
  0 & 0 & -\sin\theta
\end{array}
\right].
\end{equation}

Based on Eqs. (\ref{ele}) and (\ref{mag}), the second-order coherence properties of the scattered electric field and of the scattered magnetic field at one point $r\mathbf{s}$ in the far zone can also be found from their individual cross-spectral density matrices by setting two spatial position variables coincide, which can be given as
\begin{align}\label{polarizationmatrix1}
   \mathbf{W}^{(s,E)}(r\mathbf{s},r\mathbf{s},\omega)&=\frac{\widetilde{C}_{F}(\widehat{\mathbf{K}},\omega)}{r^{2}}\bigl[\mathbf{A}_{1}(\theta,\phi)\bigr]^{\top}\mathbf{W}(\mathbf{r}^{\prime},\mathbf{r}^{\prime},\omega)\notag \\&\times\mathbf{A}_{1}(\theta,\phi),
\end{align}
\begin{align}\label{polarizationmatrix2}
    \mathbf{W}^{(s,B)}(r\mathbf{s},r\mathbf{s},\omega)&=\frac{\widetilde{C}_{F}(\widehat{\mathbf{K}},\omega)}{r^{2}}\bigl[\mathbf{A}_{2}(\theta,\phi)\bigr]^{\top}\mathbf{W}(\mathbf{r}^{\prime},\mathbf{r}^{\prime},\omega)\notag \\&\times\mathbf{A}_{2}(\theta,\phi),
\end{align}
where $\top$ represents transpose operation of a matrix, and 
\begin{align}\label{Fouriertransform}
  \widetilde{C}_{F}(\widehat{\mathbf{K}},\omega)&=\int_{\mathcal{V}}\int_{\mathcal{V}}{C_{F}}(\widehat{\mathbf{r}}_{12}^{\prime},\omega)\exp\bigl[-i\widehat{\mathbf{r}}_{12}^{\prime \top}\widehat{\mathbf{K}}\bigr]d^{6}\widehat{r}_{12}^{\prime}
\end{align}
is the six-dimensional Fourier transform of the correlation function, with 
\begin{equation}
  {C_{F}}(\widehat{\mathbf{r}}_{12}^{\prime},\omega)=\langle F^{*}(x^{\prime}_{1},y^{\prime}_{1},z^{\prime}_{1},\omega)F(x^{\prime}_{2},y^{\prime}_{2},z^{\prime}_{2},\omega)\rangle  
\end{equation}
being the correlation function of the scattering potential of the medium (\cite{Wolf2}, Sec. 6.3.1). $\widehat{\mathbf{r}}^{\prime}_{12}=[x^{\prime}_{1},y^{\prime}_{1},z^{\prime}_{1},x^{\prime}_{2},y^{\prime}_{2},z^{\prime}_{2}]^{\top}$ and $\widehat{\mathbf{K}}=[-k(s_{x}-s_{0x}),-k(s_{y}-s_{0y}),-k(s_{z}-s_{0z}),k(s_{x}-s_{0x}),k(s_{y}-s_{0y}),k(s_{z}-s_{0z})]^{\top}$ are six-dimensional position vector and six-dimensional momentum transfer vector, respectively. 

Here we mainly focus on a wide class of random media, i.e., the so-called anisotropic, Gaussian, Schell-model medium, whose correlation function is given by the expression
\begin{align}\label{correlationfuction}
    &\noindent{C_{F}({x}^{\prime}_{1},{y}^{\prime}_{1},{z}^{\prime}_{1},{x}^{\prime}_{2},{y}^{\prime}_{2},{z}^{\prime}_{2},\omega)}\notag \\={} & C_{0}\exp{\biggl[-\frac{x^{\prime 2}_{1}+x^{\prime 2}_{2}}{4\sigma_{x}^{2}}-\frac{y^{\prime 2}_{1}+y^{\prime 2}_{2}}{4\sigma_{y}^{2}}-\frac{z^{\prime 2}_{1}+z^{\prime 2}_{2}}{4\sigma_{z}^{2}}\biggr]}\notag \\ \times{} &\exp{\biggl[-\frac{(x^{\prime}_{1}-x^{\prime}_{2})^{2}}{2\mu_{x}^{2}}-\frac{(y^{\prime}_{1}-y^{\prime}_{2})^2{}}{2\mu_{y}^{2}}-\frac{(z^{\prime}_{1}-z^{\prime}_{2})^{2}}{2\mu_{z}^{2}}\biggr]},
\end{align}
where $C_{0}$ is a positive constant. $\sigma$ and $\mu$ stands for the effective radius of the strength function and the effective radius of the normalized correlation coefficient of the medium, respectively. We now make some trivial mathematics to Eq. (\ref{correlationfuction}) and then write it in the following tensor form, viz.,
\begin{equation}\label{tensor1}
    C_{F}(\widehat{\mathbf{r}}^{\prime}_{12},\omega)=C_{0}\exp{\Bigl[-\widehat{\mathbf{r}}^{\prime \top}_{12}\mathbf{R}}\widehat{\mathbf{r}}^{\prime}_{12}\Bigr],
\end{equation}
where $\mathbf{R}$ is a six-by-six matrix, with a form of
\begin{equation}\label{tensor}
\mathbf{R}=\left[    
\begin{array}{cc}
    \mathbf{R}_{+}&  \mathbf{R}_{-}\\
     \mathbf{R}_{-}&  \mathbf{R}_{+}\\
\end{array}
\right],
\end{equation} 
where
\begin{equation}
\mathbf{R}_{\pm}=\left[
\begin{array}{ccc}
  \frac{1}{16\sigma^{2}_{x}}\pm\frac{1}{2\delta^{2}_{x}}   &  0  &\\
   0  &  \frac{1}{16\sigma^{2}_{y}}\pm\frac{1}{2\delta^{2}_{y}} & 0 \\
   0  &     0 & \frac{1}{16\sigma^{2}_{z}}\pm\frac{1}{2\delta^{2}_{z}}
\end{array}
\right]
\end{equation}
with
\begin{equation}\label{delta}
    \frac{1}{\delta^{2}_{i}}=\frac{1}{4\sigma^{2}_{i}}+\frac{1}{\mu^{2}_{i}}  \   (i=x,y,z).
\end{equation}

The momentum flow of the scattered field at any point $r\mathbf{s}$ in the far zone can be readily calculated from the Maxwell stress tensor of the field. The Maxwell stress tensor of the far field produced by scattering on a random medium is given by \cite{KG,TONG1}
\begin{align}\label{Max}
   \langle\mathbf{T}^{(s)}(r\mathbf{s},\omega)\rangle&= \frac{1}{4\pi}\Big[\mathbf{W}^{(s,E)}+\mathbf{W}^{(s,B)}\notag \\&-\frac{\mathbf{I}}{2}\text{Tr}\bigl[\mathbf{W}^{(s,E)}+\mathbf{W}^{(s,B)}\bigr]\Bigr],
\end{align}
where $\mathbf{I}$ is a three-by-three unit matrix. 

On substituting from Eq. (\ref{tensor1}) into Eq. (\ref{Fouriertransform}) first, and then into Eqs. (\ref{polarizationmatrix1}) and (\ref{polarizationmatrix2}), and finally into Eq. (\ref{Max}), after some pretty tedious calculations, the expression for the momentum flow of the far-zone scattered field, as a function of the direction of scattering $\mathbf{s}$, can be given by the expression
\begin{align}\label{mf}
  \mathbf{Q}^{(s)}(r\mathbf{s},\omega)&=\mathbf{s}\cdot\langle\mathbf{T}^{(s)}(r\mathbf{s},\omega)\rangle\notag \\ &=\frac{C_{0}\pi^{3}}{4\pi r^{2}}(\text{Det}[\mathbf{R}])^{-\frac{1}{2}}\exp{\biggl[-\frac{1}{4}\widehat{\mathbf{K}}^{\top}\mathbf{R}^{-1}\widehat{\mathbf{K}}\biggr]}\notag \\ &\times\text{Tr}\biggl[\mathbf{W}(\mathbf{r}^{\prime},\mathbf{r}^{\prime},\omega)\mathbf{Y}(\theta,\phi)\biggr]\mathbf{s},
\end{align}
where Det denotes the determinant, and 
\begin{align}
&\mathbf{Y}(\theta,\phi)={} &\notag\\&\left[
\begin{array}{ccc}
   \cos^{2}\theta\cos^{2}\phi+\sin^{2}\phi  &  -\frac{1}{2}\sin2\phi\sin^{2}\theta & -\frac{1}{2}\sin2\theta\cos\phi \\
    -\frac{1}{2}\sin2\phi\sin^{2}\theta & \cos^{2}\theta\sin^{2}\phi+\cos^{2}\phi & -\frac{1}{2}\sin2\theta\sin\phi \\ 
    -\frac{1}{2}\sin2\theta\cos\phi&-\frac{1}{2}\sin2\theta\sin\phi&\sin^{2}\theta
\end{array}
\right]
\end{align}
is a symmetric matrix.

Equation (\ref{mf}) gives the tensor form of the analytic expression for the momentum flow of the far field generated by scattering of a polychromatic electromagnetic plane wave on an anisotropic, Gaussian, Schell-model medium. From Eq. (\ref{mf}), we see that the scattered momentum flow in the far zone depends strongly on the scattering azimuthal angle $\phi$, which means that the azimuthal distribution of the scattered momentum flow in the far zone is rotationally asymmetric. Also, we see that the dependence of the scattered momentum flow on the scattering azimuthal angle $\phi$ is contained in two factors: one having to do with the exponential term, i.e., $\exp{\bigl[-\frac{1}{4}\widehat{\mathbf{K}}^{\top}\mathbf{R}^{-1}\widehat{\mathbf{K}}\bigr]}$ and the other with the trace term, i.e., $\text{Tr}\bigl[\mathbf{W}(\mathbf{r}^{\prime},\mathbf{r}^{\prime},\omega)\mathbf{Y}(\theta,\phi)\bigr]$. The former refers to the scattering medium, while the latter involves the incident source. In other words, the rotationally symmetric distribution of the scattered momentum flow in the far zone will be realized only when some constraint conditions are imposed simultaneously on the physical properties of the scattering medium and the incident field. To see these constraint conditions more clearly, we now rewrite Eq. (\ref{mf}) without tensor form,
\begin{align}\label{mf1}
     \mathbf{Q}^{(s)}(r\mathbf{s},\omega)&=H(r)\exp{\biggl[-2k^{2}\delta^{2}_{x}\sin^{2}\theta\cos^{2}\phi\biggr]}\notag \\&\times\exp{\biggl[-2k^{2}\delta^{2}_{y}\sin^{2}\theta\sin^{2}\phi\biggr]}\notag \\&\times\exp{\biggl[-2k^{2}\delta^{2}_{z}(\cos\theta-1)^{2}\biggr]} \notag \\&\times\bigg[S_{xx}(\omega)(\sin^{2}\phi+\cos^{2}\theta\cos^{2}\phi)\notag \\&-\text{Re}\bigl[S_{xy}(\omega)\bigr]\sin^{2}\theta\sin2\phi\notag \\&+S_{yy}(\omega)(\cos^{2}\phi+\cos^{2}\theta\sin^{2}\phi)\biggr]\mathbf{s},
\end{align}
where 
\begin{equation}
    H(r)=\frac{C_{0}\pi^{3}2^{\frac{9}{2}}}{4\pi r^{2}}\sigma_{x}\sigma_{y}\sigma_{z}\delta_{x}\delta_{y}\delta_{z}.
\end{equation}
Re denotes the real part, and we have used the relation $S_{yx}(\omega)=S^{*}_{xy}(\omega)$. One may easily prove $\bigl|S_{xy}\bigr|^{2}\leq S_{xx}(\omega)S_{yy}(\omega)$ in terms of the so-called Cauchy–Schwarz inequality.
\section{The necessary and sufficient conditions for producing rotationally symmetric scattered momentum flow}
Equation (\ref{mf1}) suggests that the far-zone distribution of the momentum flow of the scattered field will be independent of the scattering azimuthal angle $\phi$, i.e., that is rotationally symmetric about the direction of incidence, if the following three conditions hold simultaneously
\begin{subequations}\label{111}
\begin{align}
    \frac{1}{4\sigma^{2}_{x}}+\frac{1}{\mu^{2}_{x}}&=\frac{1}{4\sigma^{2}_{y}}+\frac{1}{\mu^{2}_{y}}\label{1}, \\
    S_{xx}(\omega)&=S_{yy}(\omega),\label{2} \\
    \text{Re}\bigl[S_{xy}(\omega)\bigr]&=0. \label{3}
\end{align}
\end{subequations}
Equation (\ref{1}) gives the constraint to the effective radius $\sigma$ and the effective correlation radius $\delta$ of the scatterer along the $x$ and $y$ axes. Although this constraint condition is formally the same as that for a planar anisoptropic source to produce a rotationally symmetric radiant intensity about the normal to the source plane \cite{LW}, one should bear this in mind that we focus on a three dimensional scatterer not a planar light source and that we calculate the momentum flow not just the radiant intensity. From Eq. (\ref{1}), one may easily see that the media have the same effective radius ($\sigma_{x},\sigma_{y}$) but different effective correlation radius ($\mu_{x},\mu_{y}$), yet all of them produce rotationally symmetric distributions of the scattered momentum flow in the far zone if Eqs. (\ref{2}) and (\ref{3}) have been met in advance. The same is true of the complementary situation, i.e., the media have the same different effective correlation radius but effective radius. In addition, Eq. (\ref{1}) can reduce to a simpler form: $\mu_{x}=\mu_{y}$ for a quasihomogeneous anisotropic meidum ($\sigma\gg\mu$) which is an important subclass of anisotropic Gaussian Schell model media \cite{DXZ, DZ}. In this case, only the effective correlation radius of the scatterer along the $x$ and $y$ axes is limited. This is because the well-known reciprocity theorem \cite{VisserTD} causes the correlation function $\widetilde{C}_{F}(\widehat{\mathbf{K}},\omega)$, and thus the scattered momentum flow to be proportional to the Fourier transform of the correlation coefficient of the scatterer. 

Equations (\ref{2}) and (\ref{3}) give the limitations to the incident source. The former requires that its spectra along the $x$ and $y$ axes must equal, whereas the latter demands that the real part of the spectral correlation between the two mutually orthogonal components $E_{x}$ and $E_{y}$ of the incident electric field must vanish. At first sight, one may be induced to conclude that an incident source which satisfies these two constraints should be completely unpolarized. However, after some simple calculations for its spectral degree of polarization, we can rule out such an inappropriate impression. The spectral degree of polarization of the optical source can be readily computed as \cite{TS}
\begin{align}\label{polarization}
    \mathcal{P}(\mathbf{r}^{\prime},\omega)=\frac{\sqrt{\bigl[S_{xx}(\omega)-S_{yy}(\omega)\bigr]^{2}+4|S_{xy}(\omega)|^{2}}}{S_{xx}(\omega)+S_{yy}(\omega)}.
\end{align}
We see that $\mathcal{P}(\mathbf{r}^{\prime},\omega)$ depends on the modulus of $S_{xy}(\omega)$, not just its real part. This fact leads that $\mathcal{P}(\mathbf{r}^{\prime},\omega)$ won't vanish even if when $\text{Re}\bigl[S_{xy}(\omega)\bigr]=0$, of course, except when $S_{xy}(\omega)$ itself is real. Therefore, Eqs. (\ref{2}) and (\ref{3}) do not impose an explicit limit on the spectral degree of polarization of the source, that is, whether the incident field is fully polarized ($\mathcal{P}(\mathbf{r}^{\prime},\omega)=1$), partially polarized ($0<\mathcal{P}(\mathbf{r}^{\prime},\omega)<1$), or completely unpolarized ($\mathcal{P}(\mathbf{r}^{\prime},\omega)=0$), it is able to a rotationally symmetric scattered momentum flow about the direction of incidence in the far zone. 

In the following, by some numerical examples, we will confirm that a rotationally symmetric scattered momentum flow in the far zone indeed can be realized provided that one  appropriately chooses the structural parameters of both the scattering medium and the incident source in terms of Eq. \eqref{111}, even though the incident electromagnetic light is scattered by an anisotropic random medium.

\begin{figure}[t]
\centering
\includegraphics[width=8cm]{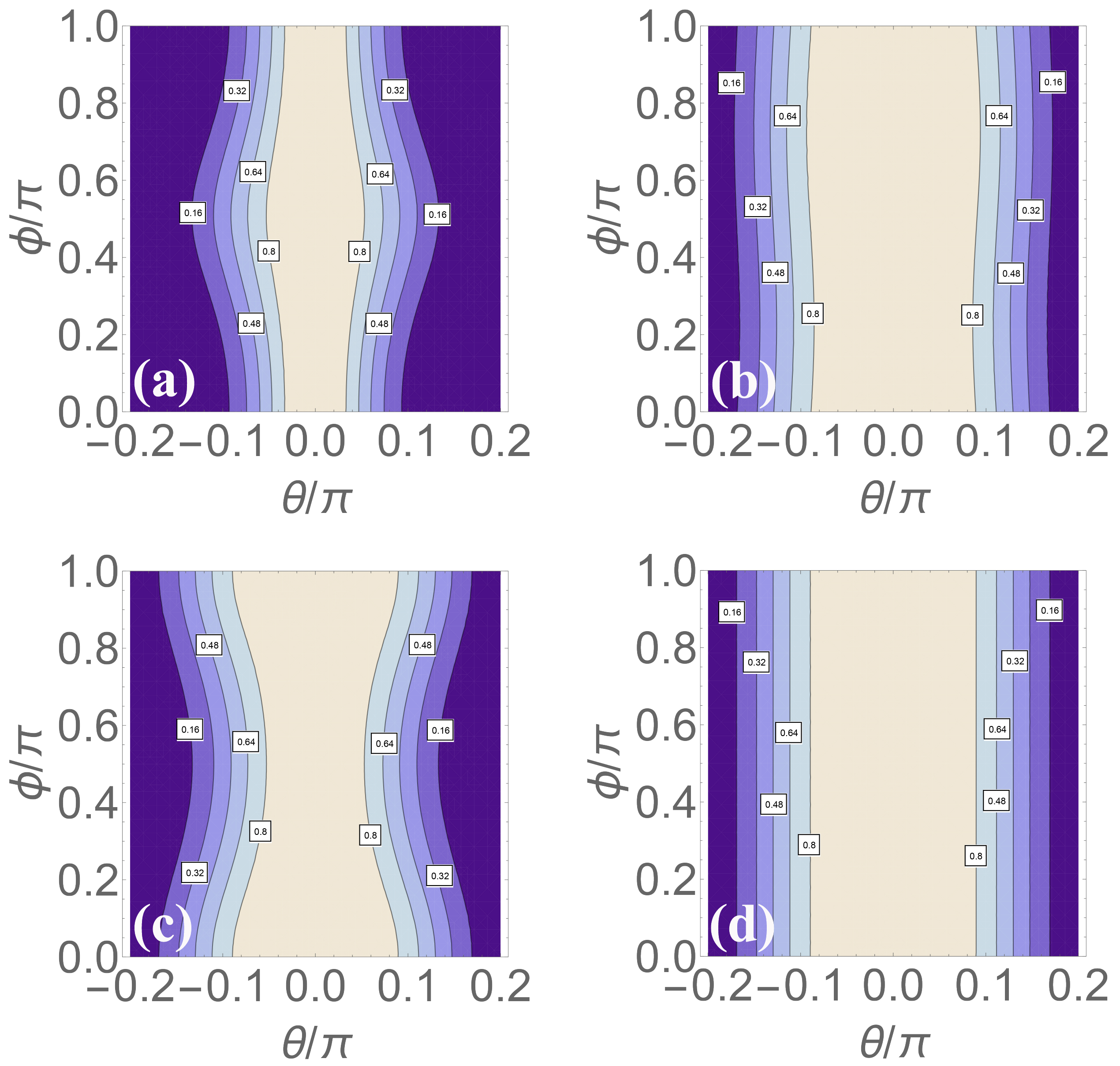}
\caption{Contours of the normalized momentum flow of the scattered field in the far zone as a function of the dimensionless polar angle $\theta/\pi$ and azimuthal angle $\phi/\pi$. $\lambda=632.8 nm$, $k=2\pi/\lambda$, $\delta_{z}=1\lambda$. The other parameters for calculations are chosen as follows: (a) $\delta_{x}=0.5\lambda$, $\delta_{y}=0.3\lambda$, $S_{xx}(\omega)=1$, $S_{yy}(\omega)=0.8$, $\text{Re}\bigl[S_{xy}(\omega)\bigr]=0.7$; (b) $\delta_{x}=\delta_{y}=0.1\lambda$, $S_{xx}(\omega)=1$, $S_{yy}(\omega)=0.8$, $\text{Re}\bigl[S_{xy}(\omega)\bigr]=0.7$; (c) $\delta_{x}=0.1\lambda$, $\delta_{y}=0.3\lambda$, $S_{xx}(\omega)=S_{yy}(\omega)=1$, $\text{Re}\bigl[S_{xy}(\omega)\bigr]=0$; (d) $\delta_{x}=\delta_{y}=0.1\lambda$, $S_{xx}(\omega)=S_{yy}(\omega)=1$, $\text{Re}\bigl[S_{xy}(\omega)\bigr]=0$.}
\label{Fig 2}
\end{figure}

Figure \ref{Fig 2} displays contours of the normalized momentum flow of the scattered field in the far zone as a function of the dimensionless polar angle $\theta/\pi$ and azimuthal angle $\phi/\pi$. Figure (\ref{Fig 2}a) plots the normalized momentum flow of the scattered field in the situation where none of three constraint conditions in Eq. (\ref{111}) holds. We see that the far-zone momentum flow of the scattering of polychromatic electromagnetic light by an anisotropic random medium is generally rotationally asymmetric about the direction of incidence. Moreover, even if either the structural parameters of the scattering medium or those of the optical source meet their individual constraint conditions in Eq. (\ref{111}), the momentum flow of the scattered field in the far zone still lacks rotational symmetry, as shown in Figs. (\ref{Fig 2}b) and (\ref{Fig 2}c). Only when each of three constraint conditions in Eq. (\ref{111}) holds, the momentum flow of the scattered field in the far zone will be roationally symmetric about the direction of incidence, as we can see from Fig. (\ref{Fig 2}d).

\section{Summary and discussion}
In summary, we have examined the far-zone momentum flow of polychromatic electromagnetic light on weak scattering from an anisotropic random medium. The tensor form of the analytic expression for the momentum flow of the scattered field in the far zone has been derived, and the far-zone distribution characteristics of the momentum flow of the scattered field have been analyzed in detail. The results have indicated that the momentum flow of plolychromatic electromagnetic light scattered by an anisotropic random medium is usually short of rotational symmetry about the direction of incidence. However, we have shown that a rotationally symmetric scattered momentum flow in the far zone would be realizable provided that the structural parameters of both the scattering medium and the incident source are selected appropriately. We have derived necessary and sufficient conditions for producing such a symmetric distribution, and demonstrated that whether the incoming source is fully polarized, partially polarized or completely unpolarized, it has the ability to produce a rotationally symmetric scattered momentum flow in the far zone. Our results not only have potential applications in the field of optical micromanipulation, but are also conducive to the inverse problem, i.e., the reconstruction of the internal information of an unknown anisotropic scatterer from measurements of the momentum flow of the scattered field in the far zone \cite{DING}. 
\section{Acknowledgements}
Financial support was provided by the National Natural Science Foundation of China No. 12204385; Natural Science Foundation of Sichuan Province No. 2022NSFSC1845; Fundamental Research Funds for the Central Universities No. 2682022CX040.

\end{document}